# PHYLOGENY-BASED TUMOR SUBCLONE IDENTIFICATION USING A BAYESIAN FEATURE ALLOCATION MODEL

By Li Zeng, Joshua L. Warren and Hongyu Zhao

*Yale University*

Tumor cells acquire different genetic alterations during the course of evolution in cancer patients. As a result of competition and selection, only a few subgroups of cells with distinct genotypes survive. These subgroups of cells are often referred to as subclones. In recent years, many statistical and computational methods have been developed to identify tumor subclones, leading to biologically significant discoveries and shedding light on tumor progression, metastasis, drug resistance and other processes. However, most existing methods are either not able to infer the phylogenetic structure among subclones, or not able to incorporate copy number variations (CNV). In this article, we propose SIFA (tumor **S**ubclone **I**dentification by **F**eature **A**llocation), a Bayesian model which takes into account both CNV and tumor phylogeny structure to infer tumor subclones. We compare the performance of SIFA with two other commonly used methods using simulation studies with varying sequencing depth, evolutionary tree size, and tree complexity. SIFA consistently yields better results in terms of *Rand Index* and cellularity estimation accuracy. The usefulness of SIFA is also demonstrated through its application to whole genome sequencing (WGS) samples from four patients in a breast cancer study.

**1. Introduction.** During cancer evolution in a patient, cancer cells acquire different genetic alterations, including single nucleotide variations (SNV[1]), copy number variations (CNV), and other more complex changes. Tumor micro-environment and treatments received by cancer patients pose selection pressure on tumor cells. As a consequence, tumor cells undergo Darwinian-like evolution, and only a few subgroups that possess better fitness survive (Nowell, 1976; Greaves and Maley, 2012; Gerlinger et al., 2012; Burrell et al., 2013). Each of these subgroups is called a subclone and has a unique genetic alteration profile. Numerous studies in different cancer types have shown that tumor subclone expansion is associated with metastasis (Ruiz et al., 2011; Yachida et al., 2010; Gerlinger et al., 2014) and drug treatment (Kreso et al., 2013; Wang et al., 2016; Ojamies et al., 2016). Identification of subclones' genotypes and reconstruction of their evolution paths can help identify possible cancer driver mutations, and bring significant insights into individualized cancer treatment (Aparicio and Caldas, 2013).

---

[1]We use SNV and mutation interchangeably in the article.

Some of the most crucial questions in the study of subclone evolution are: 1) How many subclones exist in a tumor tissue, and what are their genotypes? 2) What is the prevalence of each subclone? and 3) What is the phylogenetic structure among subclones? In recent years, researchers have devoted great efforts to answer these questions using a variety of techniques, including whole genome sequencing (WGS) (Schuh et al., 2012; Yates et al., 2015), whole exome sequencing (WES) (Carter et al., 2012), and targeted deep sequencing (Schuh et al., 2012; Campbell et al., 2008).

Figure 1 illustrates how tumor subclone information is conveyed in sequence data. The left panel in the figure displays a tumor evolutionary tree with four subclones, where the green node ($S_1$) represents normal cells and the others are cancerous subclones. We focus on three mutation loci $A, B$ and $C$. They start with normal allele status in $S_1$ (two copies of normal alleles). In subclone $S_2$, the tumor cells gain one copy of mutated allele at loci $A$ and $B$. Then $S_2$ subsequently branches to form two more subclones: $S_3$ where locus $B$ loses one normal copy, and $S_4$ where locus $C$ gains one mutated copy. Suppose we have four sequencing samples from this tumor, each being a mixture of the four subclones following mixing coefficients $\mathbf{F}$ in the middle panel. From the sequence data, we can calculate the variant allele frequency (VAF) of $A, B, C$ in each sample as shown in the right panel. The VAF for a mutation is defined as the percentage of alleles with that mutation. Note how the phylogenetic structure and subclone genotypes affect the shape of the VAF figure: mutation $C$ emerges later than $A, B$ and thus has smaller cellularity (the percentage of cells in the sequenced tumor tissue that harbor the mutation), resulting in a VAF line that lies below $A$'s and $B$'s; although mutations $A$ and $B$ emerge at the same time, $B$ later undergoes a copy loss in $S_3$, which leads to significantly higher VAF than $A$ in samples where $S_3$ takes a large fraction. With appropriate statistical modeling and inference, we can utilize the rich information hidden in tumor sequence data to study intra-tumor heterogeneity.

1.1. *Subclone inference methods.* Many popular computational tools have been developed to extract information from tumor sequence data and decipher genetic profiles for each tumor subclone.

`SciClone` (Miller et al., 2014) and `PyClone` (Roth et al., 2014) are among the earliest efforts in subclone analysis. They both identify subclones by grouping SNVs using VAF information and Bayesian clustering models. Their applications are limited to copy neutral regions, but with allele-specific copy number information available, they can also be extended to CNV regions. In addition, `SciClone` employs a variational Bayesian technique, which gives it computational advantages in large scale implementation. Instead of clustering VAF, Lee et al. (2015) proposed a non-parameteric Bayesian latent feature allocation model, `Bayclone`, to directly infer subclone SNV status. They later extended the model in (Lee et al., 2016) to handle SNVs in CNV-regions by adding a latent matrix to model CNV status of each locus.

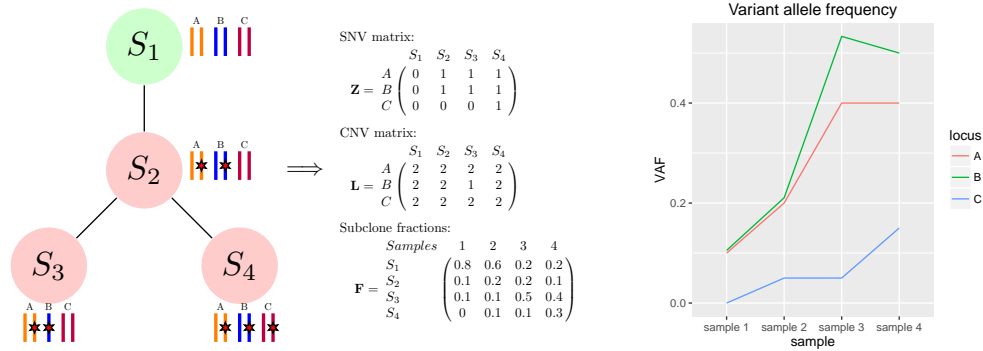

Fig 1: Overview of how sequence data inform tumor heterogeneity. The left panel displays a tumor evolutionary tree with four subclones, where the green node represents normal cells and the pink nodes are cancerous subclones. The letters $A, B$ and $C$ are mutation loci. The bars under each letter represent alleles, and the bars with red stars are mutated. The middle panel shows matrix representations of SNV and CNV status of each subclone, and $\mathbf{F}$ represents subclone fractions in each of the four sequencing samples. The right panel shows the variant allele frequency (VAF) of mutations $A, B$ and $C$ in each sequencing sample.

One major limitation of the above methods is that they fail to incorporate phylogeny in the inference which may offer insights on the temporal order of the SNVs' and CNVs' emergence. Although it is possible to mannually recover an evolutionary tree after getting subclone genotypes, such attempts will likely fail if the inference model does not explicitly enforce a phylogenetic structure. Several recent methods encompass phylogeny as a key component in subclone inference (Jiao et al., 2014; Deshwar et al., 2015; Yuan et al., 2015; Marass et al., 2017; Jiang et al., 2016). `Phylosub` (Jiao et al., 2014) is a non-parametric Bayesian model that infers subclone lineages and fractions using a tree-structured stick-breaking (TSSB) process. However it only works on copy-neutral genome regions. The later work `PhyloWGS` (Deshwar et al., 2015) extends it to CNV-regions when estimates for allele-specific copy numbers are available. `Cloe` (Marass et al., 2017) can be viewed as a tree-guided latent feature allocation model (similar to `Bayclone`). It samples posterior phylogenetic trees for each tree size within a pre-specified range, and uses model selection criterion to choose the best model. `Canopy` (Jiang et al., 2016) models tumor progression as a bifurcating tree, where SNVs and CNVs emerge along the branches and form subclones at leaf nodes. All of the phylogeny-based methods above either cannot model CNVs, or require user-provided major and minor allele number estimates, in order to incorporate copy number information.

In this article, we introduce `SIFA` (tumor **S**ubclone **I**dentification by **F**eature **A**llocation), an extension of the methods developed by Lee et al. (2016) and Marass et al. (2017), to model SNV, CNV, and phylogenetic tree under a unified framework.

**2. Model.** Subclone identification is essentially a mixture deconvolution problem. A latent phylogenetic tree and the locations where each SNV or CNV emerges jointly determine genotypes of subclones. The subclones form a mixture in tumor tissue and influence what we observe from sequence data (See Figure 1).

Assume we observe $J$ mutation loci in $T$ WGS samples from the same patient. For each locus $j$, let $d_{jt}$ denote the number of sequence reads that cover this locus, and $x_{jt}$ denote the number of reads harboring a mutation at $j$. We use $\mathbf{D}$ and $\mathbf{X}$ as the matrix form representations for the total reads and mutant reads, respectively. For notation brevity, we use $M_i$ to refer to the $i$th column of matrix $M$, and $M_{(j)}$ to denote the transpose of the $j$th row.

We use a latent integer $K$ to denote the number of subclones (normal cells included) in a tumor tissue. Then the phylogenetic tree can be represented by a vector $\mathcal{T}$ of length $K$, where $\mathcal{T}_k = i$ indicates that the parent of subclone $k$ is $i$ (if node $k$ is root, $\mathcal{T}_k = 0$). We also fix the root subclone to be normal cells (i.e. $\mathcal{T}_1 = 0$). The phylogenetic tree structure is assumed to be shared among all sequencing samples from the patient.

For modeling SNVs and CNVs, we make the following assumptions: 1) Each mutation occurs only once in a specific subclone, and is inherited by all descendant subclones (a.k.a *infinite sites assumption* (Jiao et al., 2014; Kimura, 1969)); and 2) Each CNV occurs at most once in a specific subclone, and is also inherited by descendant subclones. We are therefore able to characterize SNV and CNV status of all loci using $J \times 2$ matrices $\mathbf{Z^o}$ and $\mathbf{L^o}$, which we call *SNV* and *CNV origin matrix*, respectively. For locus $j$, $\mathbf{Z^o}_{(j)} = (k, c)$ indicates that the locus gains $c$ copies of the mutant allele in subclone $k$. Similarly, we define $\mathbf{L^o}_{(j)} = (k, c)$ to represent a gain (or loss if $c$ is negative) of $c$ copies of the normal alleles in subclone $k$. Additionally, we use $J \times K$ matrices $\mathbf{Z}$ and $\mathbf{L}$ to represent the number of mutant and total alleles for each locus in each subclone. The middle panel of Figure 1 gives examples of matrix representations of SNV and CNV status corresponding to the phylogenetic tree on its left.

WGS data also provide location information of loci. Since CNV occurs in sections of base pairs, we can harness this information to better infer CNV status. We sort the loci in the order of chromosomal positions, and divide the genome into $S$ segments, $\Delta_1, \Delta_2, \ldots, \Delta_S$. For any loci $i$ and $j$ in the same segment, we assume they share the same CNV status (i.e. $\mathbf{L^o}_{(i)} = \mathbf{L^o}_{(j)}$). In our implementation, we use R package `copynumber` (Nilsen et al., 2012) to determine genome segments. Details of this method can be found in Appendix A. Another important latent parameter is the $K \times T$ matrix $\mathbf{F}$, where the $t$th column $\mathbf{F}_t$ characterizes the fractions of each subclone in sample $t$.

Our goal is to estimate the parameters $\mathbf{Z^o}, \mathbf{L^o}, \mathbf{F}$, and the evolutionary tree $\mathcal{T}$, from the observed reads data $\mathbf{D}$ and $\mathbf{X}$. We begin by proposing a probablistic model and a Bayesian sampling algorithm for parameter inference.

2.1. *Likelihood.* In this section, we describe the statistical framework of `SIFA`. The subclone mixture, if considered as a single subclone, would have an averaged mutant allele copy number $\sum_k z_{jk} f_{kt} = \mathbf{Z}'_{(j)} \mathbf{F}_t$, and averaged total allele copy number $\sum_k l_{jk} f_{kt} = \mathbf{L}'_{(j)} \mathbf{F}_t$, for locus $j$ in sample $t$. Thus the theoretical VAF, which is also the probability for a sequenced read to be mutant if it covers locus $j$, can be calculated as the ratio of the two copy numbers:

$$(2.1) \qquad p_{jt} = \frac{\mathbf{Z}'_{(j)} \mathbf{F}_t}{\mathbf{L}'_{(j)} \mathbf{F}_t}.$$

We assume a Binomial distribution for mutant reads $x_{jt}$ conditional on total reads $d_{jt}$:

$$(2.2) \qquad x_{jt}|d_{jt}, p_{jt} \overset{ind}{\sim} \text{Binomial}(d_{jt}, p_{jt}).$$

Moreover, the number of total reads of a locus is known to be positively correlated with its averaged total copy number in WGS. We therefore use a Poisson distribution to model $d_{jt}$ (Lee et al., 2015; Klambauer et al., 2012) such that

$$(2.3) \qquad d_{jt}|\phi_t, \mathbf{L}_{(j)}, \mathbf{F}_t \overset{ind}{\sim} \text{Poisson}(\phi_t \frac{\mathbf{L}'_{(j)} \mathbf{F}_t}{2}),$$

where $\phi_t$ is the designed WGS coverage (or sequencing depth) specified when conducting sequencing. It is formally defined as the average number of times each base is sequenced. Notice that when the average total copy number $\mathbf{L}'_{(j)} \mathbf{F}_t$ is equal to 2, the distribution has mean and variance equal to $\phi_t$.

## 2.2. *Prior Specifications.*

### 2.2.1. *Clonal fractions.*
We introduce an additional parameter, $\theta_{kt}$, for each $f_{kt}$, and denote the matrix of $\theta_{kt}$'s as $\Theta$. We assign each $\theta_{kt}$ an independent Gamma$(\gamma, 1)$ prior distribution, and let

$$f_{kt} = \frac{\theta_{kt}}{\sum_{i=1}^{K} \theta_{kt}}.$$

This is equivalent to assigning $\mathbf{F}_t$ a Dirichlet$(\gamma, \gamma, \ldots, \gamma)$ prior distribution. The prior distribution has mean and mode $(1/K, 1/K, \ldots, 1/K)$, which gives no preference to any subclones. The advantage of introducing $\theta_{kt}$'s is that we can sample one $\Theta$ element at a time, while sampling $f_{kt}$ requires updating the entire vector of $\mathbf{F}_t$ due to the constraint that the $f_{kt}$ need to sum (over $k$) to one. The former approach usually leads to improved mixing using the Metropolis-Hastings sampling algorithm, thus we work with $\theta_{kt}$'s instead of $f_{kt}$'s.

### 2.2.2. *SNV and CNV origin matrices.*
For SNV, we specify a positive integer $M_S$ to be the maximum number of possible mutant copies. We utilize the following prior for $\mathbf{Z^o}$:

$$p(\mathbf{Z^o}_{(j)} = (k, c)) \propto \zeta^c \quad (\ 2 \le k \le K, 1 \le c \le M_S).$$

The above specification makes it equally likely for the SNV to originate from any subclone (except the normal subclone). Since gaining multiple copies of mutant alleles is a less likely event, our prior sets the probability of acquiring $c$ copies of the mutant alleles to be proportional to $\zeta^c$, where $\zeta$ is a pre-specified value in $(0, 1)$[2].

---

[2]We set $\zeta = 0.01$ in our implementations. Sampling results are in general robust against different choices of $\zeta$ if we do not use extremely small values, which pose strong penalty on multiple mutant copies.

Likewise, for CNV status, we specify a value $M_C$ for the maximum total allele copies. Let $(0, 0)$ represent the event of no CNV. For each genome segment $\Delta_s$, we specify the prior of its copy number status as

$$p(\mathbf{L^o}_{(j)} = (0, 0) \text{ for all } j \in \Delta_s) = \pi,$$

and uniform on other possible values. The hyper-parameter $\pi$ is given a $\text{Beta}(a_\pi, b_\pi)$ prior.

2.2.3. *Phylogenetic tree.* We have previously introduced a $K$-vector representation $\mathcal{T}$ of the tree structure, where the root node is fixed to be normal cells. We add one additional constraint: for any $2 \leq k \leq K$, we require that $\mathcal{T}_k \in \{1, 2, \ldots, k-1\}$. Note that under this constraint $\mathcal{T}_2$ can only be 1, which implies that the second subclone is a direct descendant of normal cells. It can be shown that any vector $\mathcal{T}$ satisfying these constraints is a tree, and for any tree of size $K$, we can find a corresponding representation satisfying the constraints (see Appendix B). We use a discrete uniform distribution on all possible tree structures as the prior distribution.

Figure 2 presents the parameters' dependencies in the `SIFA` model. The dark grey nodes represent the observed data, while the light grey nodes represent the latent parameters. Our ultimate goal is to make inference about $\Theta, \mathbf{L^o}, \mathbf{Z^o}$, and $\mathcal{T}$ from the posterior distribution $p(\Theta, \mathbf{L^o}, \mathbf{Z^o}, \mathcal{T}, \pi | \mathbf{D}, \mathbf{X}, \Phi)$, where $\Phi = (\phi_1, \phi_2, \ldots, \phi_T)$ represents the read depths of the sequencing samples.

2.3. *Posterior Sampling.* For brevity of description, we let $\Omega$ denote the set of all unknown parameters, and $\Omega_{-\omega}$ denote all unknown parameters except $\omega$.

We employ Gibbs sampling to acquire posterior samples, and therefore need to sample from the full conditional distributions for each parameter. However, since the full conditional distributions for some parameters are in forms from which direct sampling is difficult, we also employ the Metropolis-Hastings sampling in such cases.

2.3.1. *Subclone fractions.* Subclone fractions can be fully represented by $\Theta$. We update one entry of $\Theta$ at a time. Since the full conditional distribution $p(\theta_{kt} | \mathbf{D}, \mathbf{X}, \Phi, \Omega_{-\theta_{kt}})$ cannot be directly sampled, we employ the Metropolis-Hastings sampling at this step.

Let $\theta_{kt}$ be the current sample in our Markov chain. A new value $\theta_{kt}^*$ is proposed from the transition function $h(\theta_{kt}^* | \theta_{kt}, s)$, where $h(\cdot | \theta_{kt}, s)$ is the density function for distribution $\text{Gamma}(s\theta_{kt}, s)$, which is centered at $\theta_{kt}$ and has variance $\theta_{kt}/s$. The tuning parameter $s$ controls the proposing step-length, and a larger $s$ value usually leads to a higher acceptance rate. In our implementation, we adaptively tune its value to keep acceptance rate in a reasonable range[3] in order to ensure efficient mixing of the Markov chain.

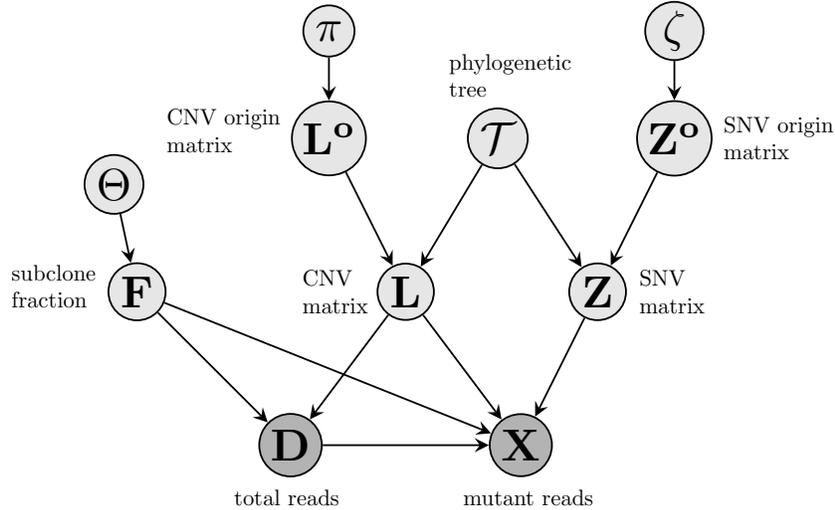

Fig 2: Illustration of the parameters' dependency. The dark grey nodes in the figure stand for the observed data, and the light grey ones represent the latent parameters.

2.3.2. *SNV, CNV origin matrices.* Since the sampling spaces for CNV status $\mathbf{L^o}$, SNV status $\mathbf{Z^o}$ are discrete and relatively small, we can perform Gibbs sampling by calculating the probability for every possible status.

Because loci are independent when sampling SNV status, we update $\mathbf{Z^o}$ row by row. For each $j$, we calculate $p(\mathbf{Z^o}_{(j)} = (k,c)|\mathbf{D}, \mathbf{X}, \Phi, \Omega_{-\mathbf{Z^o}_{(j)}})$ for all possible combinations of $(k,c)$, and use them as weights to sample a new $\mathbf{Z^o}_{(j)}$. Sampling of $\mathbf{L^o}$ follows a similar procedure, but instead of updating one locus at a time, one segment of loci is updated together.

We also apply Gibbs sampling to $\mathbf{L^o}$'s hyper-parameter $\pi$, because its full conditional distribution directly takes the form of a Beta distribution

$$p(\pi|\mathbf{L^o}) \sim \text{Beta}(n + a_\pi, S - n + b_\pi),$$

where $S$ is the number of genome segments and $n$ is the number of segments

---

[3]We used range $[0.4, 0.65]$ in our implementation.

without CNV.

2.3.3. *Phylogenetic tree.* The sample space for the phylogenetic tree $\mathcal{T}$ is discrete as well, including all possible tree structures of size $K$. The size of this space is however much larger than $\mathbf{L^o}$'s and $\mathbf{Z^o}$'s, and grows fast as $K$ increases. It is computationally impractical to calculate all possible outcome probabilities in this case. Instead, we propose a mixed sampling approach, where we randomly choose to perform the Metropolis-Hastings sampling or slice sampling (Neal, 2003).

For the Metropolis-Hasting sampling, we employ a simple proposal method: randomly select a leaf node and rewire it to a randomly selected parent node. However, this kind of proposal is likely to be rejected, since changes in tree structure have strong impact on model likelihood. To illustrate this, we create a toy example phylogenetic tree with four subclones in the left panel of Figure 3. The capital letters $A, B, C, D$ are SNVs or CNVs, and the subclones they reside in are their origin subclones. We use the lowercase letters $a, b, c, d$ to represent subclone fractions. If we change the tree structure by rewiring node $D$ from its original parent $B$ to a new parent $C$, then the cellularity of $C$ increases from $c$ to $c + d$, while cellularity for $B$ decreases from $b + d$ to $b$. If many loci are involved in $B$ and $C$, the proposal will lead to dramatic change in the full conditional likelihood. Since the parameters have been sampled in favor of the original tree structure, the change in likelihood is most likely in the decreasing direction, which will lead to poor acceptance rate.

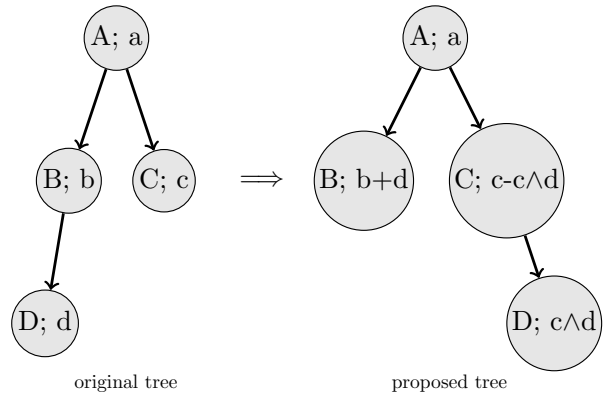

original tree                 proposed tree

Fig 3: Example of the Metropolis-Hastings tree proposal. Node $D$ is the randomly selected tree leave to be rewired.

To address this issue, we introduce the following procedure:

1. Randomly select a leaf subclone $k$ (whose current parent is $p$), and a valid target parent subclone $q$. Rewire $k$ to $q$ to get the new tree $\mathcal{T}^*$.

2. Change the previous parent's fractions:

$$\theta_{pt}^* = \theta_{pt} + \theta_{kt} \quad (t = 1, 2, \ldots, T)$$

3. Change the new parent's fractions:

$$\theta_{qt}^* = \theta_{qt} - \theta_{kt} \wedge \theta_{qt} \quad (t = 1, 2, \ldots, T)$$

4. Assign new fractions to subclone $k$:

$$\theta_{kt}^* = \theta_{kt} \wedge \theta_{qt} \quad (t = 1, 2, \ldots, T)$$

where $x \wedge y$ stands for the minimum of $x$ and $y$. Then $\mathcal{T}^*$ and $\Theta^*$ can be used to calculate acceptance probability and perform sampling. It can be shown that, under the condition that $\theta_{qt} \geq \theta_{kt}$ for all $t$, the proposed tree has exactly the same likelihood as the original tree. The proof is provided in Appendix C. Following this sampling strategy, the proposed tree in the previous example takes the form as displayed in the right panel of Figure 3. Notice that only the cellularity for $D$ is changed, while $B$ and $C$ remain intact.

Although the above Metropolis-Hastings algorithm is tuned to achieve an improved acceptance rate, current trees can only change to trees differing by one subclone. It is possible that the algorithm will be stuck at a locally optimal tree structure. Therefore, we also include the slice sampling technique (see Appendix D) as an option in the tree sampling scheme. In each iteration, we

1. Sample a nuissance parameter $u^* \sim \text{Uniform}(0, p(\mathcal{T}|\mathbf{D}, \mathbf{X}, \Phi, \Omega_{-\mathcal{T}}))$
2. Randomly and repeatedly propose $\mathcal{T}^*$ from all possible tree structures and accept the proposal if $p(\mathcal{T}^*|\mathbf{D}, \mathbf{X}, \Phi, \Omega_{-\mathcal{T}}) \geq u^*$

Slice sampling enables our sampler to make bigger jumps and avoids getting trapped at local modes. In our empirical analysis, a combination of slice sampling and the Metropolis-Hastings sampling increased our sampler's mobility, yielded improved sampling performance, and was robust in different simulation scenarios.

Derivations of the full conditional distributions for all model parameters are provided in Appendix E.

2.3.4. *Parallel tempering.* In MCMC sampling, the technique of parallel tempering is often adopted to make it easier to jump from one mode to another (Gelman et al., 2014).

We specify an increasing sequence of temperatures $\{t_1, t_2, t_3, \ldots, t_c\}$, usually with equally spaced $t_i = 1 + \Delta T(i-1)$, where $\Delta T$ is the temperature increment. For each temperature, we specify a corresponding target function $p_i(\Omega) = p(\Omega | \mathbf{D}, \mathbf{X}, \Phi)^{1/t_i}$. The $c$ target functions share the same set of modes, however, functions with higher temperature will look more "flat" which makes it easier for Bayes samplers to explore all modes. We run one independent chain for each target. At user-specified intervals, we randomly select a chain $i < c$ and propose to switch chains $i$ and $i + 1$. This proposal is accepted with the following probability (Geyer, 1991):

$$\frac{p_i(\Omega^{(j)})p_j(\Omega^{(i)})}{p_i(\Omega^{(i)})p_j(\Omega^{(j)})},$$

where $\Omega^{(k)}$ is the latest sample from chain $k$. In the end, samples from chain 1 are taken as posterior samples of $p(\Omega | \mathbf{D}, \mathbf{X}, \Phi)$.

2.4. *Model selection.* After obtaining posterior samples from models with different numbers of subclones, we need to address the issue of model selection. Obviously, choosing the model with the maximum likelihood will lead to overfitting, because models with more subclones are more likely to yield an improved likelihood. Researchers have utilized a variety of model selection methods for subclone inference. Marass et al. (2017) used log-posterior likelihood of the Maximum a posteriori (MAP) sample as the selection criterion. However, this approach largely depends on the choice of prior distributions. If prior distributions are flat, selection by posterior distribution and by maximum likelihood are equivalent. Many methods (Jiang et al., 2016; Parisi et al., 2011; Li and Li, 2014; Zare et al., 2014) adopt the Bayesian information criterion (BIC), which adds a penalty term to the negative log-likelihood in order to penalize complex models. However, there is also concern as how to calculate the number of free model parameters, since the model involves both discrete (SNV/CNV status and tree structure) and continuous (subclone fractions) parameters, and they should not contribute equally to model complexity.

We choose to use a criterion based on Bayes free energy, which is defined as

$$\mathcal{F} = -\log \int p(\mathbf{X}, \mathbf{D} | \Omega, \Phi)\psi(\Omega) \, d\Omega,$$

where $\psi(\Omega)$ is the joint prior density of all parameters. $\mathcal{F}$ can be understood as the negative logarithm of the marginal likelihood of a model. Selecting the model with the largest marginal likelihood is equivalent to choosing the model that minimizes the Bayes free energy. It has been established that, in regular statistical models, BIC asymptotically approximates $\mathcal{F}$. However, this is not generally true when the underlying model is singular (e.g. when posterior distribution have multiple modes) (Watanabe, 2013). A computational way to evaluate $\mathcal{F}$ is

$$(2.4) \qquad \hat{\mathcal{F}} = -\sum_{j=1}^{c} \log \mathbf{E}_{\Omega}^{\beta_{j+1}} \left[ \exp(-(\beta_j - \beta_{j+1})\mathcal{L}(\Omega)) \right],$$

where $1 = \beta_1 > \beta_2 > \ldots > \beta_{c+1} = 0$ are the inverse temperatures (i.e. $\beta_j = 1/t_j$ for $1 \le j \le c$), and $\mathcal{L}(\Omega) = -\log(\mathbf{X}, \mathbf{D}|\Omega, \Phi)$ is the negative log-likelihood. The empirical expectation $\mathbf{E}_{\Omega}^{\beta_j}$ is calculated using MCMC samples from the $j$th chain(Watanabe, 2013).

**3. Simulation.** In order to assess the performance of `SIFA`, we first generated simulation datasets under different sequencing depths, true subclone numbers, and tree structures, and compared inference results with `Pyclone` (Roth et al., 2014), one of the most popular intra-tumor heterogeneity analysis methods, and `Cloe` (Marass et al., 2017), one of the most recently published methods.

3.1. *Simulation setup.* We compared the methods in eleven simulation scenarios in total, including $3 \times 3$ basic simulation scenarios: three values of sequencing depth $= 40, 60, 80$ reads per base-pair and three values of true subclone number $K = 3, 4, 5$. We used tree structure $(0, 1, 1)$ for $K = 3$, $(0, 1, 2, 2)$ for $K = 4$, and $(0, 1, 1, 2, 2)$ for $K = 5$. We used the same set of $\mathbf{L}$ and $\mathbf{Z}$ matrices for the same $K$ (see Figure 4). The sequencing depths 40 and 60 are commonly adopted in WGS practices, and depth 80 is considered as high coverage. In all simulations, we fixed the number of loci $J = 200$, and the number of sequencing samples $T = 4$, which also resemble real data from intra-tumor heterogeneity studies.

In order to study the robustness of the methods against different tree structure specifications, we considered two other scenarios at $K = 5$, depth $= 40$ reads per base-pair. Note that the tree $(0, 1, 1, 2, 2)$ for $K = 5$ in the basic scenarios had three leaf nodes. In order to experiment on trees with different complexities, we further generated two more tree structures: $(0, 1, 2, 2, 3)$ with two leaf nodes, and $(0, 1, 2, 3, 4)$ with one leaf node. More details about simulation setup can be found in Appendix F.

3.2. *Measure of model performance.* We evaluated the performance of our method using different metrics, including the ability to recover true phylogenetic trees, the ability to recover the correct number of subclones, mutation clustering accuracy, and mutation cellularity estimation accuracy. Since `Pyclone` does not perform inference for the phylogenetic tree, we only used the last two criteria to compare `SIFA` with `Pyclone` and `Cloe`.

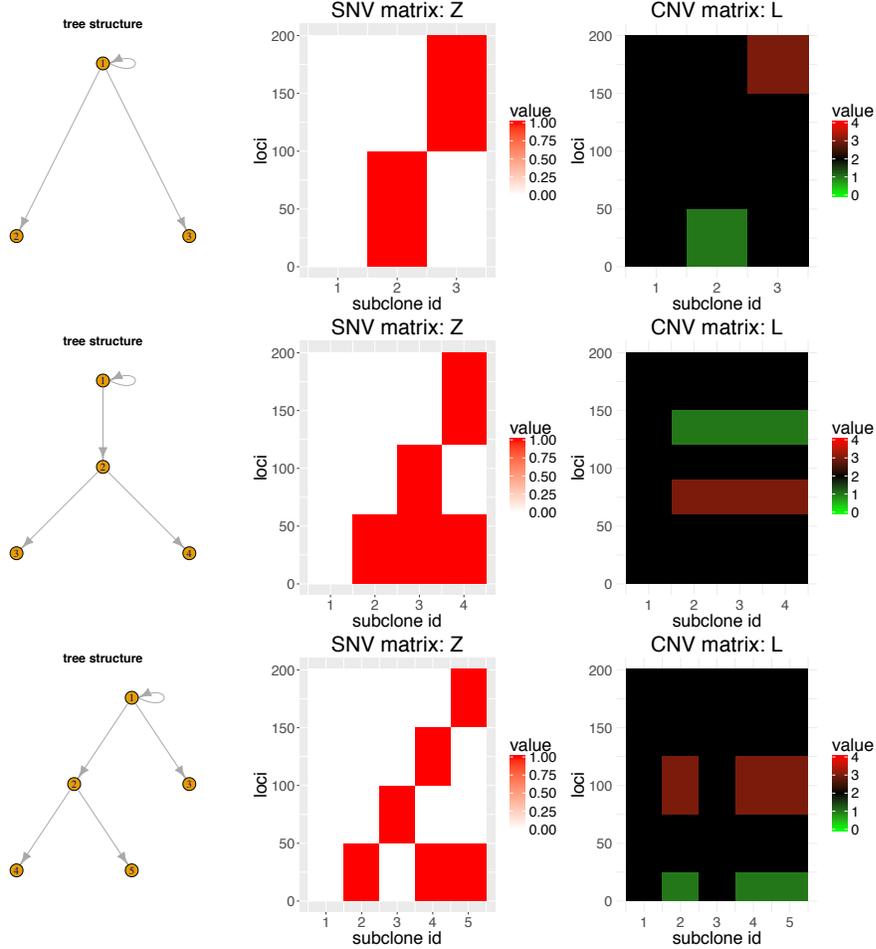

Fig 4: Simulation parameters. The three rows represent simulation parameters (including tree structure $\mathcal{T}$, SNV matrix $\mathbf{Z}$, and CNV matrix $\mathbf{L}$) for $K = 3, 4, 5$, respectively. The fist column shows the underlying phylogenetic trees. The second column presents SNV matrices, where the red blocks represent one mutant copy gain. The third column presents CNV matrices, where the green blocks represent one copy loss, the red blocks represent one copy gain, and the black blocks are copy neutral.

The ability to recover true tree structures and subclone numbers can be assessed by directly comparing model selection results with true $\mathcal{T}$ and $K$. For mutation clustering accuracy, we used the *Rand Index* (Rand, 1971), which measures the similarity between two partitions. Clustering is basically a partitioning of the SNV set $A = \{SNV_1, SNV_2, \ldots, SNV_J\}$, into non-overlapping groups. Let $P^{(1)} = \{P_1^{(1)}, P_2^{(1)}, \ldots, P_r^{(1)}\}$ and $P^{(2)} = \{P_1^{(2)}, P_2^{(2)}, \ldots, P_s^{(2)}\}$ be two partitions that divide $A$ into $r$ and $s$ groups, respectively. The *Rand Index* looks at all pairs of $(SNV_i, SNV_j)$ and count how many pairs are assigned to the same group. It is formally defined as:

$$R(P^{(1)}, P^{(2)}) = \frac{TP + TN}{\binom{J}{2}},$$

where

1. $TP$ (true positive) : pairs of SNVs that are in the same group in $P^{(1)}$ and also in the same group in $P^{(2)}$
2. $TN$ (true negative) : pairs of SNVs that are in different groups in $P^{(1)}$ and also in different groups in $P^{(2)}$.

We calculated this index using posterior partitions acquired from our MCMC sampler and the true partitions. The *Rand Index* takes values in $[0, 1]$, with larger value indicating higher clustering accuracy. The maximum value of 1 is achieved when the two partitions are exactly the same.

Calculation of cellularity estimation error is straight-forward. Given SNV and subclone fraction matrix, cellularity of $SNV_j$ in sample $t$ is $c_{jt} = \sum_{k \in \Gamma_j} f_{kt}$, where $\Gamma_j = \{k : z_{jk} > 0\}$ is the set of subclones that bear mutation $j$. Let $C$ be the $J \times T$ true cellularity matrix and $\hat{C}$ be the cellularity estimation calculated from a single draw from the posterior distribution. Cellularity estimation error is defined as the mean of absolute element-wise difference between $C$ and $\hat{C}$:

$$C_{err} = \frac{\sum_{j,t} |c_{jt} - \hat{c}_{jt}|}{JT}.$$

3.3. *Simulation results.* When running `SIFA`, we used the first 2000 iterations to tune the adaptive parameter $s$ which controls the step size for proposals in sampling $\Theta$ (see section 2.3.1), 4000 iterations for burn-ins, and the next 4000 samples for inference. We also ran `Pyclone` and `Cloe` for 10,000 iterations and kept the last 4000 samples for inference. When implementing `Pyclone`, we provided the true minor and major copy numbers for each locus. Convergence of the MCMC samples was tested using the Geweke

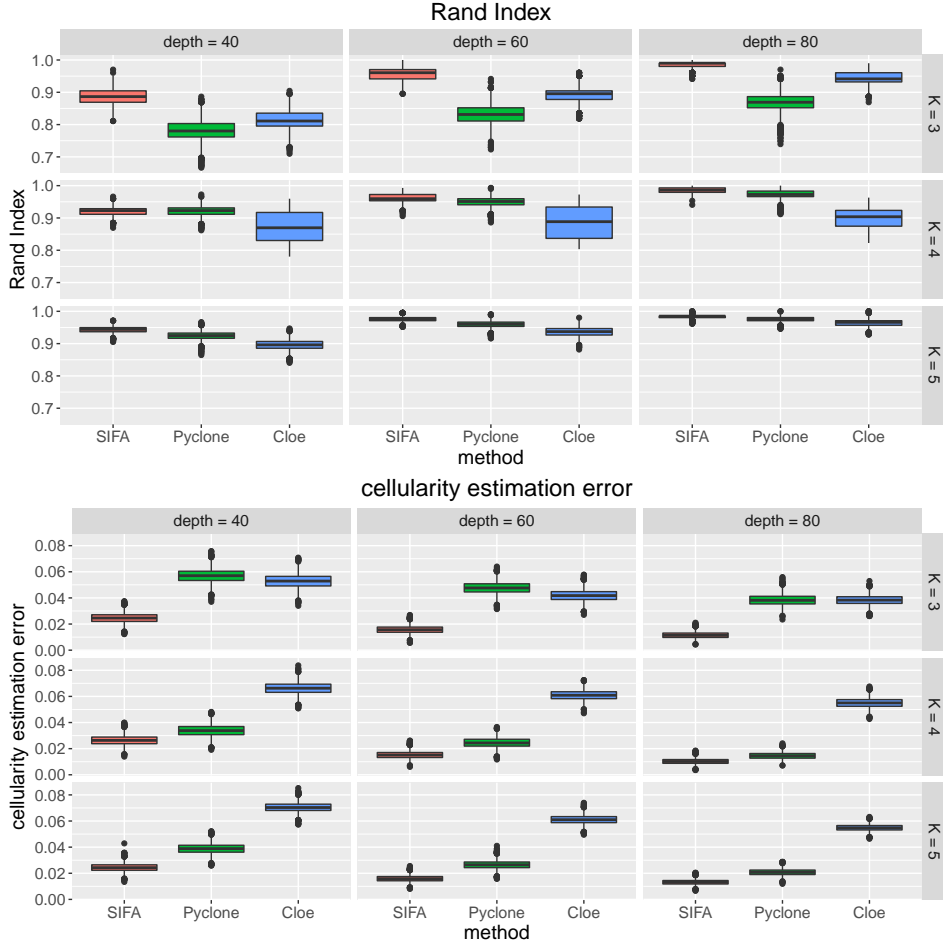

Fig 5: Comparisons of method performance using the *Rand Index* and cellularity estimation error ($C_{err}$). The upper panel and lower panel respectively show results for *Rand Index* and $C_{err}$ in the nine basic simulation scenarios. Rows represent scenarios with different numbers of underlying subclones, and columns represent different sequencing depths. Within each scenario, the three boxes display posterior distributions of the two criteria calculated using posterior samples from `SIFA`, `Pyclone`, and `Cloe`, respectively.

statistic (Brooks and Gelman, 1998) (see supplementary materials, Figure 29).

`SIFA` successfully recovered the true phylogenetic structures in all eleven simulation scenarios. Figure 5 presents the comparison between `SIFA`, `Pyclone` and `Cloe`, in terms of *Rand Index* and $C_{err}$. Results for the two additional scenarios are shown in Appendix F. Rows in Figure 5 represent scenarios with different numbers of underlying subclones, and columns represent different sequencing depths. Within each scenario, the three boxes display the posterior distributions of the two criteria calculated using the posterior samples from each method. As expected, for each fixed $K$, all three methods yielded better performances as simulation sequencing depth increased from 40 to 80. Under scenario $K = 5$, depth $= 40$ and tree structure 3, `SIFA` did not outperform `Pyclone`, because one CNV segment was not correctly identified, leading to incorrect mutation assignments in subclones 2 and 3 (see Figure 4 of the supplementary material and Appendix F). In all other settings, `SIFA` had the best results in both *Rand Index* and cellularity estimation. `Pyclone` also yielded smaller error than `Cloe` in most cases, but note that `Pyclone` took the ground truth minor and major copy number as input, which gave it certain prior advantages in comparison.

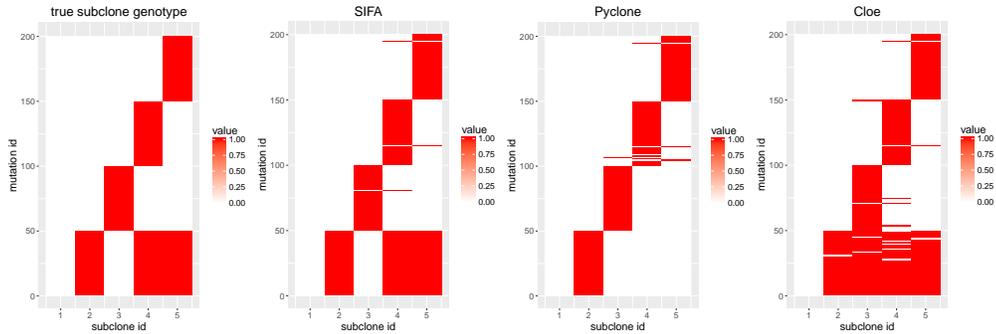

Fig 6: Point estimates of the SNV matrix $\mathbf{Z}$ from the three methods under scenario $K = 5$, depth $= 60$. The first panel represents the gold-standard $\mathbf{Z}$ matrix, and the other three are point estimates from `SIFA`, `Pyclone` and `Cloe`, respectively.

We also plot the ground truth $\mathbf{Z}$ matrix and its point estimates from the three methods in Figure 6 (take scenario $K = 5$ and depth $= 60$ as example). For `SIFA` and `Cloe`, point estimates were calculated using the posterior median from a selected $K$. Since `Pyclone` yielded varying $K$ in posterior samples, we manually picked the $K$ with the largest posterior frequency, and calculated the median $\mathbf{Z}$ matrix using the corresponding subset of samples. Full details on calculating the posterior point estimates are described in Appendix H. From Figure 6, we can see that all three methods can correctly detect the overall clustering pattern. `Pylcone` did not perform phylogeny inference, thus the subclones have no overlapping SNVs. `Cloe` correctly inferred genotypes of four out of five subclones, except subclone 2 whose parent should be the normal subclone.

More detailed results for each simulation are available in Figures 1 - 8 in the supplementary materials.

## 4. Application: Breast Cancer.

In this section, we apply our method to a breast cancer dataset (Yates et al., 2015). This study involved 50 breast cancer patients, where researchers applied WGS and targeted sequencing to multiple samples from each of the 50 patients' tumors. We applied `SIFA` to patients PD9694, PD9771, PD9777 and PD9849, who have greater or equal to three WGS samples.

To obtain genome segmentation estimates, we sorted the patients' loci by their chromosomal locations, and ran the `multipcf` segment calling method using the loci's total reads information. In WGS, the loci may give false mutated reads due to technical errors in the sequencing process. In order to screen out "noise" loci and only keep those that were shared among samples, we performed a few steps of quality control using criteria including locus coverage and average VAF across samples (see Appendix G). An overview of the final input data is presented in Table 1.

TABLE 1
*Overview of the breast cancer dataset*

| patients | number of loci | number of segments |
|----------|----------------|--------------------|
| PD9694 | 301 | 56 |
| PD9771 | 549 | 152 |
| PD9777 | 742 | 83 |
| PD9849 | 641 | 56 |

We applied `SIFA` with $K = 3, 4, \ldots, 7$ to the four patients' sequence data, and selected the models that minimized $\mathcal{F}_n$. We present the results for patients PD9694 and PD9777, and leave the others' results in the supplementary materials. The `SIFA` parameters we used for the implementations can be found in Appendix G.

Patient PD9694 developed multifocal breast cancer. Three WGS samples were acquired from the patient, with designed sequencing depths 41, 48 and 43, respectively. Two out of three samples (PD9694a and PD9694c) were from invasive foci, and PD9694d was from an area of DCIS (ductal carcinoma in situ), which was non-invasive.

Results for PD9694 are presented in the upper row of Figure 7. Our model selection criterion recommended $K = 5$ as the best model. The left panel displays the inferred phylogenetic tree structure and corresponding genes for each subclone. The genes presented are among the 184 most mutated driver genes in breast cancer as reported by the IntOGen-mutation platform (Gonzalez-Perez et al., 2013). We use different colors to indicate copy number status: green for copy loss, red for copy gain, and black for copy neutral genes. Gene SF3B1 is involved in RNA splicing process, and has been found to be a hotspot mutation gene in breast cancer. Knowing that it originated from an early phase subclone, we could use targeted therapy for treatment, such as Spliceostatin A, which is reported to be effective on SF3B1 mutated cell lines (Maguire et al., 2015). The middle panel presents the composition of the subclones in each sample. The major components of the DCIS sample (PD9694d) are subclones 2 and 3, and both were formed in the early stage of evolution. The two invasive samples each developed a new major subclone (subclone 5 for PD9694c and subclone 4 for PD9694a). It is possible to identify gene alterations related to tumor invasiveness from these subclones.

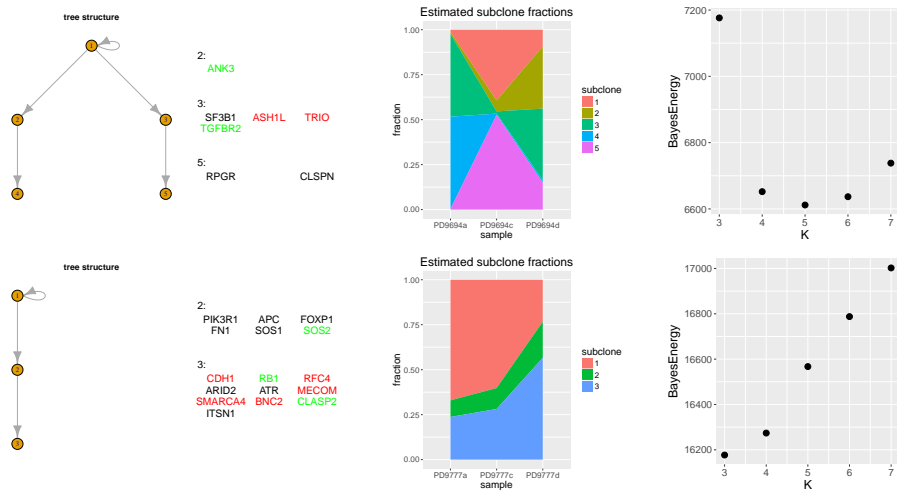

Fig 7: Breast cancer analysis results for PD9694 (first row) and PD9777 (second row). The left panels present the inferred phylogenetic tree, with breast cancer related genes listed on its right. Red gene names indicate loci with copy gain, green names indicate loci with copy loss, and the others are copy neutral loci. The middle panels show fractions of each subclone in all WGS samples. Subclones are represented by colors, and the lengths of each colored segment are proportional to their estimated fractions. The right panels present the Bayes free energy values calculated in the model selection step.

The second row of Figure 7 presents the results for a triple negative breast cancer patient PD9777. The three WGS samples, PD9777a, PD9777c and PD9777d, have designed sequencing depths 30, 30 and 57, respectively. Sample PD9777d was collected from the tumor tissue after neo-adjuvant chemotherapy, and the other two were from pre-treatment tumor mass. The fraction plot in the middle panel depicts a contraction of the normal subclone and an expansion of a cancerous subclone (subclone 3) in the post-therapy sample, indicating the tumor has developed chemotherapy-resistance. Gene RB1 might explain the resistance and the expansion of subclone 3 in PD9777d. It is a negative regulator of the cell cycle and is the first tumor suppressor gene found. It has also been reported to be associated with drug sensitivity in triple negative breast cancer (Robinson et al., 2013). Another interesting finding is that, three out of six driver genes (PIK3R1, SOS1, SOS2) originated from subclone 2 can be mapped back to the PIK3 pathway, suggesting that this pathway was severely mutated in the early stage of tumor development. PIK3 is a major intracellular signaling pathway, and there are rich literatures studying its association with cell growth and tumor proliferation. This finding provides strong evidence that, the PIK3 pathway mutations played an important role in the tumor progression of PD9777.

CNV estimation is generally more reliable for larger segments than smaller ones. When a segment is short, there is a higher probability that the observed reads of the segment are higher (or lower) than average (the designed sequencing coverage) just by chance, and subsequently this leads to false positives in CNV inference. For example, a 3-loci-segment showing below average reads may happen randomly, but a 50-loci-segment showing below average reads is a strong indicator of copy loss. To control for false positive CNV calls, we set the prior distribution of $\pi$ (the prior probability of a segment being copy neutral) close to one, thus a CNV is not assigned unless it provides considerable increase in model likelihood.

The inferred CNVs of the previous two patients are demonstrated in Figure 8. The Y-axis shows the log-transformed normalized reads, and segments without CNVs should contain points centered around zero. The orange points in the figure are loci without estimated CNV, while other colors indicate CNVs on different chromosomes. From Figure 8, we observe that most of the segments that have consistent deviations from zero are colored, which manifests the ability of SIFA to successfully capture the major CNV regions. In patient PD9694, a copy loss was detected in a segment that covers gene PTEN, a known tumor suppressor gene on chromosome 10. The copy loss took place in subclone five, which is a major component of sample PD9694c, and can possibly be the cause of the invasiveness of this sample. In PD9849, we detected strong evidence of CNV at a mutated gene FGFR2 on chromosome 10, whose observed reads were 2-4 times above average (see Figure 14 of the supplementary materials). It belongs to the Fibroblast Growth Factor Receptor family, and is found to be a consistent top hit in genome-wide association studies of breast cancer(Campbell et al., 2016). `SIFA` inference suggested that multiple mutant FGFR2 alleles were acquired at an early stage of the tumor, which could be a cause of tumorigenesis. More details on the CNV inferences for PD9771 and PD9849 are presented in Figures 13 - 14 of the supplementary materials.

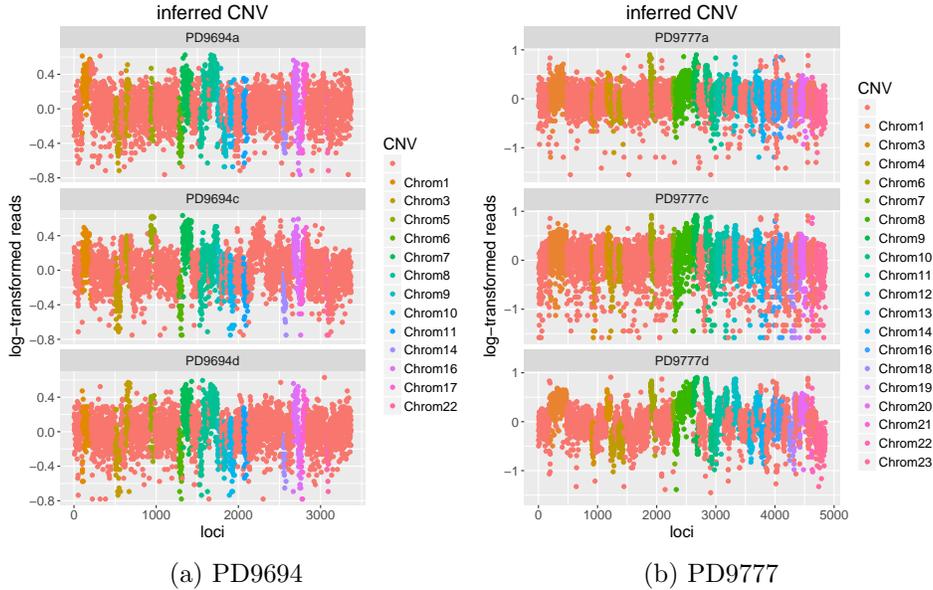

(a) PD9694                                    (b) PD9777

Fig 8: CNVs detected by SIFA for PD9694 and PD9777. The X-axis represents mutated loci arranged by their chromosomal locations. The Y-axis represents the log-transformed loci coverage: the reads for each locus are divided by median reads and then log-transformed. Loci with no estimated CNV are indicated with orange points, while points in other colors indicate CNVs on different chromosomes.

We also applied `Cloe` and `Pyclone` to the dataset, and compared their inferences with the results from `SIFA`[4]. `SIFA` and `Cloe` yielded similar overall SNV clustering patterns (see supplementary material Figure 21-26). `Pyclone` reported significantly more clusters than the other two methods for patient PD9694, but many clusters only had smaller than five mutations, which might be due to overfitting noises. Moreover, `SIFA`'s estimates led to better VAF fitting as shown in Figure 27 of the supplementary material. We calculated a VAF matrix estimation from each posterior sample in all three methods, computed its mean absolute difference to the observed VAF matrix, and the distribution of the differences are presented as the boxes in the figure. In all patients, `SIFA` has significantly lower VAF fitting error, indicating that the incorporation of CNV information and phylogeny structure can indeed improve model performance.

---

[4] `Pyclone` was only applied to patients PD9694 and PD9849, since it needed estimated major and minor allele copy numbers as input. This information was missing for the other two patients.

**5. Discussion and Conclusion.** In recent years, many studies have shown that intra-tumor heterogeneity plays an important role in tumor metastasis and treatment resistance. It is of great clinical importance to correctly infer patients' subclonal composition and phylogenetic structure. For this purpose, we proposed `SIFA`, which is an extension of (Marass et al., 2017) and (Lee et al., 2015) that enables simultaneous inference of subclone SNV, CNV status and phylogenetic tree. `SIFA` achieved satisfactory results in our simulation studies, and its application to four patients from a breast cancer study also provided interesting insights.

In the Metropolis-Hastings sampling step of tree structures in `SIFA`, we employed a novel tree proposal method which includes changing the tree structure and the subclone fraction parameters at the same time. The method successfully led to improved acceptance rate and better mixing efficiency. We also provided proof that, under certain constraints, the proposed tree has equivalent likelihood as the original one. Note that the constraints can also serve as conditions for unidentifiability: if the underlying true fractions for subclone $D$ are indeed smaller than $C$ in all samples (see Figure 3), then the original tree and the proposed tree are indistinguishable in terms of likelihood. In this case, our method is able to produce both trees in posterior samples, and the users can incorporate other prior information to decide which tree structure is more likely to be true.

Our model runs one chain for each candidate $K$ and uses the Bayes free energy as the model selection criterion. BIC is used as model selection criterion by some other methods. Although BIC is simpler in form, it requires users to provide the number of free model parameters. Calculating this number is not trivial as we cannot simply count the number of parameters in the model, when there are both continuous ($\mathbf{F}$) and discrete ($\mathbf{L}, \mathbf{Z}, \mathcal{T}$) parameters that do not contribute equally to model complexity. Moreover, the Bayes free energy criterion utilizes all posterior samples to assess model pre-dictive accuracy, which should give more comprehensive results compared to criteria using pointwise estimates (such as AIC, BIC and DIC), which are known to have issues when the posterior distributions have multiple modes.

In SIFA, we make several assumptions about the tumor evolution process, which may limit the application in real data. For example, the infinite sites assumption states that a locus can only be mutated once. While this is true for most mutation loci in practice, users may encounter several loci with multiple mutation types (e.g. $A \to C$ and $A \to G$). Such cases indicate violation of the infinite sites assumption, and we suggest users exclude such mutation loci from the analysis. Second, when modeling the subclone fractions parameter $\Theta$, we assume the samples are independent, and therefore can update $\Theta$ column by column. However, considering that the samples are sometimes from the same tissue, dependence among samples may exist. Especially when the biopsy sites are close to each other, the subclone fractions may appear to be similar as well. In such cases, modeling the dependency among samples will likely provide improved inference of subclone fractions. Third, SIFA is only suitable for analysis of WGS data. It may not be applicable to samples from other sequencing technologies, such as WES and targeted deep sequencing, because the Poisson distribution assumption in Equation 2.3 may not hold. Under such circumstances, other analysis tools should be considered.

In all applications, we have used maximum number of mutated alleles $M_S = 2$ and maximum copy number $M_C = 4$. The choices of the two numbers may seem low for modeling extreme cases where loci acquire many copies of mutated alleles or are subject to strong CNV, but it should be flexible enough to model most mutation loci. Since the subclone inference is dependent on mutation clusters, if most loci are correctly modeled, the inference results should be robust against a few extreme cases. To further examine the model's sensitivity against different choices of $M_S$ and $M_C$, we fit SIFA on the real dataset with $M_S = 3$ and $M_C = 6$. The resulting estimation for $\mathbf{Z}$ and $\mathbf{L}$ were similar to the results reported in Section 4, other than a few loci that were estimated to have larger numbers of mutated copies or total copies. Inference for the subclone phylogeny also remained the same for three of the four patients, except for patient PD9771 where the new tree had a $(0, 1, 2, 2)$ structure. However, this tree was also present in the posterior samples from previous analysis as a minority tree.

Theoretically, SIFA can handle sampling for models with any $K$. However in practice, the sampler becomes more time consuming as $K$ gets larger. It can take an unnecessarily long time if a large range of $K$'s is used. It is suggested that users explore smaller ranges first and check the model selection metric. If the metric suggests the possibility of larger number of subclones, for example when you see the metric keeps decreasing as K increases, then it is necessary to explore a few more values for $K$.

Another limitation of `SIFA` is that users need to run the model for each $K$ in the pre-specified range. It may be preferable to spend less time on less likely $K$s. One possible future direction is to enable the sampler to jump between models with different $K$s, for example reconstructing the model with a non-parametric prior for $K$. This modification is likely to increase sampling efficiency, and at the same time bypass the need of model selection.

Source code and instructions for implementation of `SIFA` can be accessed from **https://github.com/zengliX/SIFApackage**.

## APPENDIX A: SEGMENTATION METHOD

For genome segmentation, we use the `multipcf` function in R package `copynumber` developed by Nilsen et al. (2012). It is a penalized regression model with penalty placed on the number of segments. Suppose we observe normalized copy number measurements $\mathbf{y_i} = (y_1^i, y_2^i, \ldots, y_n^i)$ for the $i$th sample, where $i = 1, 2, \ldots, T$. Let $S = \{I_1, I_2, \ldots, I_m\}$ be a genome segmentation, where each $I_k$ contains indices of loci in the $k$th segment. In each segment, the loci are supposed to take the same copy number value, thus `multipcf` minimizes the following loss function over $S$:

$$L(S|\mathbf{y_1}, \mathbf{y_2}, \ldots, \mathbf{y_T}, \gamma) = \sum_{i=1}^{T} \sum_{I \in S} \sum_{k \in I} (y_k^i - \bar{y}_I{}^i)^2 + \gamma |S|$$

where $\bar{y}_I{}^i$ is the mean copy number of probes in segment $I$ of sample $i$, $|S|$ is the number of segments, and $\gamma$ is a penalty parameter. Larger values of $\gamma$ indicate a larger penalty for opening a new segment. In our implementation, we select the optimal $\gamma$ using BIC.

## APPENDIX B: TREE REPRESENTATION PROOF

In this section, we will prove that any tree of size $K$ can be represented as a $K$-vector satisfying the constraints in Section 2.2.3, and any such vector corresponds to a tree.

We focus on cases when $K \geq 2$.

Given any tree, we sort the nodes in decreasing order of their heights, and index them by $1, 2, \ldots, K$. We then create the vector $\mathcal{T}$ by assigning $\mathcal{T}_k$ to be the index of node $k$'s parent. Apparently index 1 refers to the root node, thus $\mathcal{T}_1 = 0$. And for any node $k \geq 2$, its parent must have greater height, thus $\mathcal{T}_k \in \{1, 2, \ldots, k-1\}$ holds true.

Given any $K$-vector $\mathcal{T}$ satisfying the constraints, we let the first entry be root node, and add an edge between nodes $i, k$ if $\mathcal{T}_k = i$. Apparently any node $k \geq 2$ has a path to the root, which impies the graph is connected. The graph also has only $K - 1$ edges, thus it must be a tree.

## APPENDIX C: EQUIVALENT TREE PROPOSAL PROOF

In this section, we prove that, under the condition that $\theta_{qt} \geq \theta_{kt}$ for all $t$, the new tree proposed in our Metropolis-Hastings sampling step yields the same likelihood as the original tree structure.

It is easy to check that the proposed $\Theta^*$ leads to the following changes in the fraction matrix $\mathbf{F}$:

$$f_{pt}^* = f_{pt} + f_{kt}$$

(C.1)
$$f_{qt}^* = f_{qt} - f_{kt} \wedge f_{qt}$$

(C.2)
$$f_{kt}^* = f_{kt} \wedge f_{qt} \qquad (\text{for all } t = 1, 2, \ldots, T)$$

Figure 9 illustrates the tree structures before (left panel) and after (right panel) rewiring subclone $k$ to subclone $q$. To prove both tree structures yield the same likelihood, we need to show that the theoretical VAF $p_{jt} = \sum_k z_{jk} f_{kt} / \sum_k l_{jk} f_{kt}$ stays the same for every locus. It suffices to prove that for every SNV and CNV, its cellularity does not change in the new tree. Since the proposal only directly affects subclones $k, p$ and $q$, we can focus our proof just in these three subclones.

When the condition $\theta_{qt} \geq \theta_{kt}$ holds, $f_{kt} \wedge f_{qt}$ becomes $f_{kt}$. Thus equations C.1 and C.2 can be simplified to $f_{qt}^* = f_{qt} - f_{kt}$ and $f_{kt}^* = f_{kt}$.

- For any SNV or CNV $g_k$ in subclone $k$, it has cellularity $f_{kt}$ in both trees.
- For any $g_p$ in subclone $p$, it has original cellularity $f_{pt} + f_{kt} + c_p$, where $c_p$ is the sum fraction of $p$'s other descendant sucblones. In the new tree, the value becomes $f_{pt}^* + c_p = f_{pt} + f_{kt} + c_p$, thus the invariance holds.
- For any $g_q$ in subclone $q$, its original cellularity is $f_{qt} + c_q$. And cellularity in the new tree is $f_{qt}^* + f_{kt}^* + c_q = f_{qt} + c_q$, where $c_q$ is similarly defined as $c_p$. Cellularity invariance also holds true.

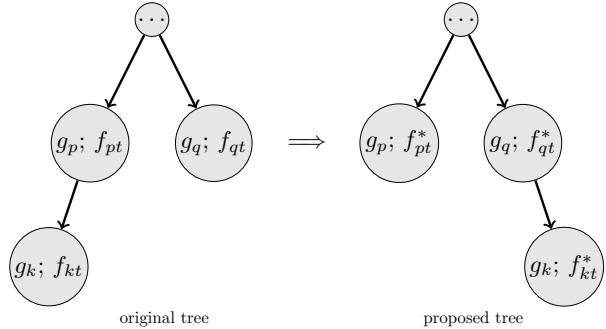

Fig 9: Left panel: original tree; right panel: proposed tree

## APPENDIX D: SLICE SAMPLING

The method of slice sampling is commonly used to sample distributions with bounded domain. It enables the sampler to jump between different local modes even if they are far from each other, thus avoiding getting biased samples.

Suppose our goal is to sample $z$ from distrbution $f(z) = Cg(z)$ and we only know $g(z)$, which is usually the case in Bayesian models. We introduce a nuisance random variable $u|z \sim \text{Unif}(0, g(z))$, then $u, z$ have joint distribution

$$p(z, u) = C\mathbf{1}_{\{\mathbf{u} \leq \mathbf{g(z)}\}}$$

We intend to get samples of $f(z)$ by sampling from the joint distribution $p(z, u)$ using Gibbs Sampling. In each iteration, we

1. Sample $u^* \sim \text{Unif}(0, g(z))$
2. Sample $z^* \sim p(z|u^*) \propto \mathbf{1}_{\{g(z) \geq u^*\}}$, which is uniform on area $\{z : g(z) \geq u^*\}$. Therefore, we can randomly propose $z^*$ from its support and accept proposal if $g(z^*) \geq u^*$

One practical issue in computation is that $g(z)$ is usually near 0 when the sample size is large, which makes it infeasible to sample from $\text{Unif}(0, g(z))$. However, this can be fixed by working on the log scale. Let $\eta = -\log(u)$, then $\eta$ has a shifted Exponential distribution: $\text{Exp}(1) - \log g(z)$, which is much easier to sample from.

Therefore, in the actual implementation of slice sampling, we use the following more practical procedure in each iteration:

1. Sample $\eta^* \sim \text{Exp}(1) - \log g(z)$
2. Randomly propose $z^*$ from its support and accept proposal if $\log g(z^*) \geq -\eta^*$

## APPENDIX E: DERIVATIONS OF THE FULL CONDITIONED DISTRIBUTIONS

• Posterior distribution of $\pi$

Posterior distribution for $\pi$ is (with prior $\text{Beta}(a_\pi, b_\pi)$):

$$
\begin{aligned}
p(\pi|\mathbf{L^o}) &\propto p(\mathbf{L^o}|\pi)p(\pi) \\
&\propto \pi^n(1-\pi)^{S-n}p(\pi) \\
&\propto \text{Beta}(n + a_\pi, S - n + b_\pi)
\end{aligned}
$$

where $S$ is the number of genome segments and $n$ is the number of segments without CNV.

● Posterior distribution of $\Theta$

Since samples are independent, we can update $\Theta$ by columns.

$$
\begin{aligned}
p(\Theta_t|\mathbf{D}, \mathbf{X}, \Phi, \Omega_{-\Theta_t}) &= p(\Theta_t|\phi_t, D_t, \mathbf{L^o}, \mathbf{Z^o}, \mathcal{T}) \\
&\propto p(\mathbf{D}_t|\phi_t, \Theta_t, \mathbf{L}) p(\mathbf{X}_t|\mathbf{D}_t, \mathbf{Z}, \mathbf{L}, \Theta_t) p(\Theta_t) \\
&\propto \prod_j (\mathbf{Z}'_{(j)} \Theta_t)^{x_{jt}} [(\mathbf{L}_{(j)} - \mathbf{Z}_{(j)})' \Theta_t]^{d_{jt}-x_{jt}} e^{-\frac{\phi_t}{2G_t} \mathbf{L}'_{(j)} \Theta_t} p(\Theta_t)
\end{aligned}
$$

where $G_t = \sum_k \theta_{kt}$.

● Posterior distribution of $\mathbf{Z^o}$

$\mathbf{Z^o}$ is sampled row by row:

$$
\begin{aligned}
p(\mathbf{Z^o}_{(j)}|\mathbf{D}, \mathbf{X}, \Phi, \Omega_{-\mathbf{Z^o}_{(j)}}) &= p(\mathbf{Z^o}_{(j)}|\mathbf{X}, \mathbf{D}, \Theta, \mathbf{L^o}, \mathcal{T}) \\
&\propto p(\mathbf{Z^o}_{(j)}) \prod_t p(x_{jt}|d_{jt}, p_{jt}) \\
&\propto p(\mathbf{Z^o}_{(j)}) \prod_t p_{jt}^{x_{jt}} (1-p_{jt})^{d_{jt}-x_{jt}} \\
&\propto p(\mathbf{Z^o}_{(j)}) \prod_t (\mathbf{Z}'_{(j)} \mathbf{F}_t)^{x_{jt}} (\mathbf{L}'_{(j)} \mathbf{F}_t - \mathbf{Z}'_{(j)} \mathbf{F}_t)^{d_{jt}-x_{jt}}
\end{aligned}
$$

● Posterior distribution of $\mathbf{L^o}$

Rows of $\mathbf{L^o}$ in the same segment are sampled together. Consider loci in segment $\Delta_i$, and assume they share the same CNV status $\mathbf{L^o}_{(\Delta_i)}$:

$$
\begin{aligned}
p(\mathbf{L^o}_{(\Delta_i)}|\mathbf{D}, \mathbf{X}, \quad &\Phi \quad, \Omega_{-\mathbf{L^o}_{(j:j\in\Delta_i)}}) = p(\mathbf{L^o}_{(\Delta_i)}|\mathbf{D}, \mathbf{X}, \Phi, \Theta, \mathbf{Z^o}, \mathcal{T}) \\
&\propto p(\mathbf{L^o}_{(\Delta_i)}) \prod_{j\in\Delta_i} \prod_t p(d_{jt}|\phi_t, \mathbf{F}_t, \mathbf{L^o}_{(j)} = \mathbf{L^o}_{(\Delta_i)}, \mathcal{T}) \times \\
&\qquad \prod_{j\in\Delta_i} \prod_t p(x_{jt}|d_{jt}, \mathbf{F}_t, \mathbf{Z^o}_{(j)}, \mathbf{L^o}_{(j)} = \mathbf{L^o}_{(\Delta_i)}, \mathcal{T}) \\
&\propto p(\mathbf{L^o}_{(\Delta_i)}) \prod_{j\in\Delta_i} \prod_t [(\mathbf{L}_{(\Delta_i)} - \mathbf{Z}_{(j)})' \mathbf{F}_t]^{d_{jt}-x_{jt}} e^{-\frac{\phi_t}{2} \mathbf{L}'_{(\Delta_i)} \mathbf{F}_t}
\end{aligned}
$$

where $\mathbf{L}_{(\Delta_i)}$ is the length-$K$ subclone copy number vector corresponding to $\mathbf{L^o}_{(\Delta_i)}$ when tree structure is $\mathcal{T}$.

● Posterior distribution of $\mathcal{T}$

$$
\begin{aligned}
p(\mathcal{T}|\mathbf{D}, \mathbf{X}, \Phi, \Omega_{-\mathcal{T}}) &= p(\mathcal{T}|\mathbf{D}, \mathbf{X}, \Phi, \mathbf{L^o}, \mathbf{Z^o}, \Theta) \\
&\propto p(\mathcal{T}) \prod_{j,t} p(d_{jt}|\phi_t, \Theta_t, \mathbf{L^o}, \mathcal{T}) p(x_{jt}|d_{jt}, \Theta_t, \mathbf{L^o}, \mathbf{Z^o}, \mathcal{T})
\end{aligned}
$$

## APPENDIX F: SIMULATION ANALYSIS

**F.1. Simulation true parameters.**    Details of the simulation parameters are presented in Figure 10.

**F.2. Parameter specifications for Bayesian sampling.**    Model parameter specifications are presented in Table 2. Markov chain sampling parameters are presented in Table 3.

TABLE 2

*Prior distribution parameters used in the applications of SIFA to simulations*

| Parameters | value |
|---|---|
| number of samples $(T)$ | 4 |
| number of loci $(J)$ | 200 |
| maximum possible mutant copy $(M_S)$ | 2 |
| maximum possible total copy $(M_C)$ | 4 |
| $\Theta$ Dirichlet prior parameter $(\gamma)$ | 1.5 |
| $\pi$ Beta prior parameter ( $a_\pi, b_\pi$ ) | (10000,1) |
| $\mathbf{Z^o}$ prior parameter $(\zeta)$ | 10 |

TABLE 3

*MCMC sampling parameters used in the applications of SIFA to simulations*

| Parameters | value |
|---|---|
| Number of chains | 8 |
| Temperature increment $(\Delta T)$ | 0.35 |
| posterior sample size | 4000 |
| burn-in samples size | 4000 |
| sample size for adaptive parameter tuning | 2000 |
| interval to perform chain swap | 30 |
| probability to perform tree slice sampling | 0.15 |
| probability to perform tree Metropolis-Hastings sampling | 0.85 |

**F.3. Simulation results on different tree structures.**    Results are presented in Figure 11.

## APPENDIX G: REAL DATA ANALYSIS

**G.1. Data quality control.**    There were more than $3000 \sim$ SNV loci in the raw data for different patients, and many of them were present due to sequencing noises. To screen out loci that were less useful for our analysis and reduce input sample size, we went through the following process:

1. Removed loci with total reads $\leq 15$ in any samples
2. Calculated observed VAF, and removed loci with average VAF $\leq 0.1$

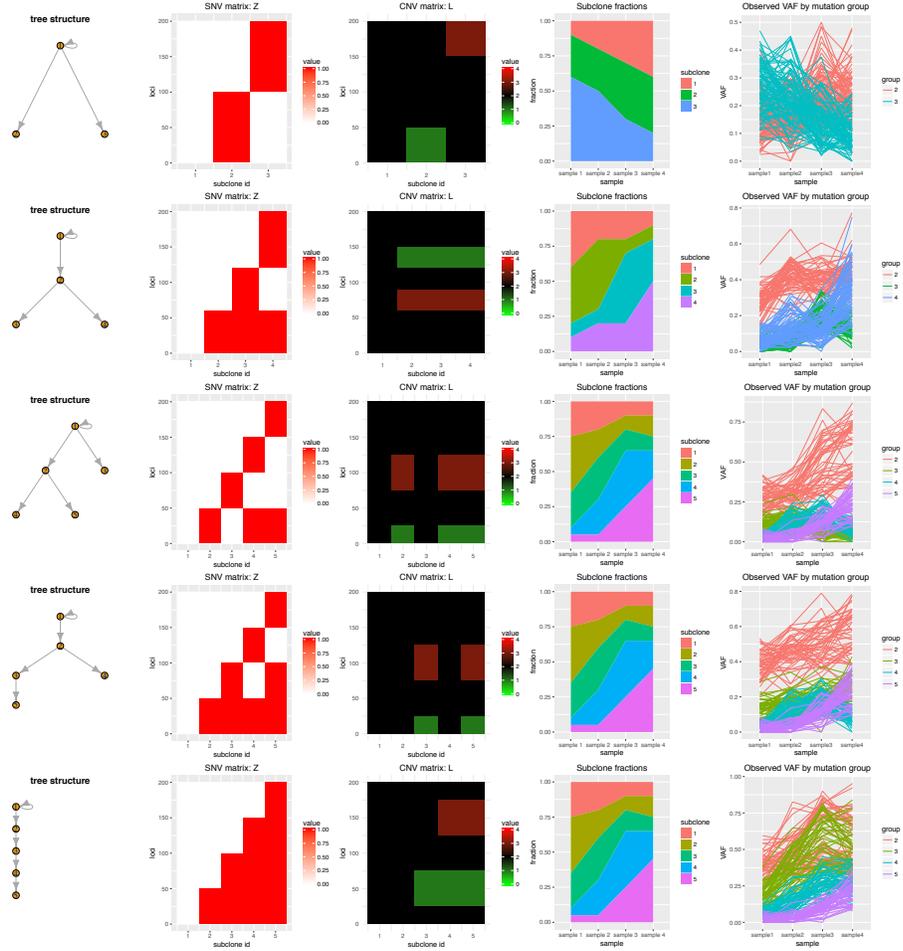

Fig 10: The underlying truth for simulations. The first two rows display simulation parameters for $K = 3, 4$, and the last three rows for $K = 5$. The second column is SNV matrix, where the red blocks represent one mutant copy gain. The third column is CNV matrix, where the green blocks represent one copy loss, the red blocks indicate one copy gain, and the black blocks are copy neutral. The fourth column depicts subclone fraction fluctuations across samples, where the vertical lengths of the colored segments represent subclones' fractions. The fifth column is the observed VAF (one line for each locus) generated from corresponding parameters in each row (at sequencing depth 40). Mutations originated from different subclones are coded with different colors.

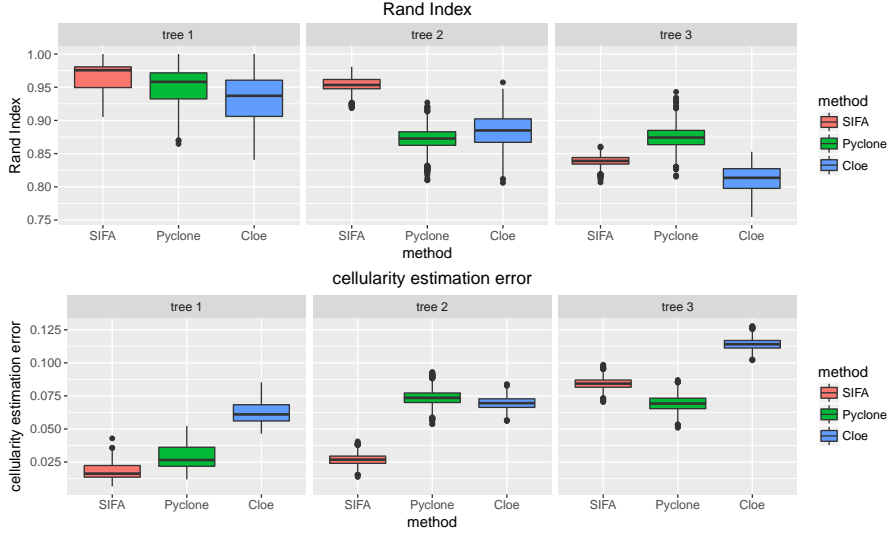

Fig 11: Simulation results under scenario $K = 5$, depth $= 40$ and different tree structures.

3. Performed K-means clustering using loci's VAF with specified number of clusters = 60. In each cluster, the loci had highly similar VAF patterns, thus we randomly removed half of the loci to reduce redundant information

**G.2. Sampling parameters.** Model parameter specifications for real data analysis are presented in Table 4. Markov chain sampling parameters are presented in Table 5.

TABLE 4

*Prior distribution parameters used in the applications of SIFA to the breast cancer dataset*

| Parameters | value |
|---|---|
| number of samples ($T$) | 4 |
| number of loci ($J$) | 200 |
| maximum possible mutant copy ($M_S$) | 2 |
| maximum possible total copy ($M_C$) | 4 |
| $\Theta$ Dirichlet prior parameter ($\gamma$) | 1.5 |
| $\pi$ Beta prior parameter ( $a_\pi, b_\pi$ ) | (10000,1) |
| $\mathbf{Z}^\circ$ prior parameter ($\zeta$) | 0.01 |

## APPENDIX H: CALCULATION OF POINT ESTIMATES

It is not trivial to obtain posterior point estimates, because under different tree structures the parameters $\mathbf{Z}, \mathbf{L}$, and $\mathbf{F}$ have different interpretations. In this section, we describe in detail how the point estimates are calculated.

Note that one phylogenetic tree may have different representations. For example, if a tree has two leaf subclones with the same parent, we can switch them in the tree and their corresponding columns/rows in $\mathbf{Z}, \mathbf{L}$, and $\mathbf{F}$, and yield an equivalent representation of the tree. To unify posterior samples with different tree parameters that may refer to the same tree, we employ the following procedure to process each posterior sample:

1. Let $\mathcal{T}, \mathbf{Z}$ be the tree and SNV matrix estimate in the current sample, and similarly define $\mathcal{T}_{prev}, \mathbf{Z}_{prev}$ for its previous sample;
2. If $\mathcal{T} \neq \mathcal{T}_{prev}$, we list all permutations (of columns) of $\mathbf{Z}$ and identify the permutation $\sigma$ that minimizes the average absolute difference between $\mathbf{Z}$ and $\mathbf{Z}_{prev}$;
3. Reindex $\mathcal{T}$ by permutation $\sigma$. If the resulting new tree is valid (satisfies the constraints specified in Section 2.2.3), we also change $\mathbf{Z}, \mathbf{L}$, and $\mathbf{F}$ according to $\sigma$. Otherwise we make no change to this sample.

After the above postprocessing, we typically end up with a single tree structure. Then we use median $\mathbf{Z}, \mathbf{L}$, and mean $\mathbf{F}$ as point estimates. However, it is possible to have more than one tree in the posterior samples. It is not reasonable to combine samples from essentially different trees to calculate point estimates. In such cases, we report the point estimates from the majority tree, but it is also suggested to examine all the tree structures in the samples.

TABLE 5
*MCMC sampling parameters used in the applications of SIFA to the breast cancer dataset*

| Parameters | value |
| --- | --- |
| Number of chains | 8 |
| Temperature increment ($\Delta T$) | 0.35 |
| posterior sample size | 4000 |
| burn-in samples size | 4000 |
| sample size for adaptive parameter tuning | 2000 |
| interval to perform chain swap | 30 |
| probability to perform tree slice sampling | 0.15 |
| probability to perform tree the Metropolis-Hastings sampling | 0.85 |

## ACKNOWLEDGEMENTS

This publication was supported by the NIH grants P50 CA196530, P01 CA154295, and CTSA Grant Number UL1 TR001863 from the National Center for Advancing Translational Science (NCATS), a component of the National Institutes of Health (NIH).

We would also like to thank Jiehuan Sun for discussions during model development, and Xiaoxuan Cai for her advice on this article.

## SUPPLEMENTARY MATERIAL

**Supplement A: supplementary figures; replication program**
(https://github.com/zengliX/SIFApackage/tree/master/Supplementary).

Li Zeng
Joshua L. Warren
60 College Street
New Haven, CT 06510
E-mail: li.zeng@yale.edu
E-mail: joshua.warren@yale.edu

Hongyu Zhao
60 College Street, Ste 201
New Haven, CT 06510
E-mail: hongyu.zhao@yale.edu